\documentclass[sigconf,nonacm]{acmart}
 
\AtBeginDocument{%
  }

\usepackage{subcaption}
\usepackage{booktabs}
\usepackage[figuresright]{rotating}
\usepackage{array}
\usepackage{multirow}
\usepackage[utf8]{inputenc}
\usepackage{tabularx}
\usepackage{float}
\usepackage{balance}
\usepackage{footmisc}

\usepackage{calc}

\usepackage{graphicx}
\usepackage{longtable}
\usepackage{tabularx}
\usepackage{adjustbox}
\usepackage{rotating}
\usepackage{afterpage}
\usepackage{pifont}
\usepackage{amsfonts}
\usepackage{flafter}
\usepackage{xcolor}
\usepackage[most]{tcolorbox}
\usepackage{pdflscape}
\usepackage{adjustbox}
\usepackage{ragged2e}
\usepackage{balance}
\usepackage{kvoptions}
\usepackage{fancybox}
\usepackage{setspace}

\makeatletter
\SetupKeyvalOptions{family=PB, prefix=PB@}
\DeclareStringOption[promptbody]{placeholder}
\DeclareStringOption[prompttitle]{title}
\DeclareStringOption[blue!50!black]{bordercolor}
\DeclareStringOption[blue!2!white]{fillcolor}
\DeclareStringOption[blue!70!black]{titlecolor}
\DeclareStringOption[1.5pt]{linewidth}
\DeclareBoolOption[true]{showtitle}
\ProcessKeyvalOptions{PB}

\newcommand{\promptbox}[1][]{%
  \setkeys{PB}{#1}%
  \colorlet{pb@border}{\PB@bordercolor}%
  \colorlet{pb@fill}{\PB@fillcolor}%
  \colorlet{pb@title}{\PB@titlecolor}%
  \noindent
  \ifPB@showtitle
    \fboxrule=\PB@linewidth
    \fboxsep=0pt
    \fcolorbox{pb@border}{pb@border}{%
      \hspace{4pt}%
      \textcolor{white}{\small\bfseries\PB@title}%
      \hspace{4pt}%
      \vphantom{Ag}%
    }%
    \vspace{-\fboxrule}%
    \par\noindent
  \fi
  \fboxrule=\PB@linewidth
  \fboxsep=8pt
  \fcolorbox{pb@border}{pb@fill}{%
    \begin{minipage}{\dimexpr\linewidth-2\fboxrule-2\fboxsep-2pt}%
      \vspace{4pt}%
      \textcolor{black}{\small\PB@placeholder}%
      \vspace{6pt}%
    \end{minipage}%
  }%
  \par
}
\makeatother

\begin{document}

\title{CARE: A Capability-Based Measurement Framework for Reproductive Equity in Human-AI Interaction}

\author{Alice Zhong}
\authornote{Equal contributions.}
\affiliation{%
  \institution{Society-Centered AI Lab \\ UNC Chapel Hill, USA}
  \country{}
}\email{azhong@unc.edu}

\author{Phoebe Chen}
\authornotemark[1]
\affiliation{%
  \institution{Society-Centered AI Lab \\ UNC Chapel Hill, USA}
\country{}}\email{pkc@unc.edu}  

\author{Punya Aragula}
\affiliation{%
  \institution{School of Data \& Information Sciences, UNC Chapel Hill, USA}
  \country{}
}\email{punyara@unc.edu}

\author{Anika Sharma}
\affiliation{%
  \institution{Society-Centered AI Lab \\UNC Chapel Hill, USA}
  \country{}}
\email{anika17@unc.edu}

\author{Kandyce Brennan}
\affiliation{%
  \institution{School of Nursing \\ UNC Chapel Hill, USA}
  \country{}
}\email{heddrick@unc.edu}

\author{Snehalkumar~`Neil'~S.~Gaikwad}
\correspondingauthor
\affiliation{%
  \institution{Society-Centered AI Lab \\ UNC Chapel Hill, USA}
  \country{}}
\email{gaikwad@cs.unc.edu}

\renewcommand{\shortauthors}{Zhong,  Chen, Sharma, Aragula, Brennan, and Gaikwad}


\begin{abstract}
Algorithmic systems mediate sexual and reproductive health (SRH) information seeking. Standard HCI and AI evaluation centers usability, accuracy, and interaction quality, measures designed to assess task performance and interaction quality at the system level. We introduce CARE, the Capability Approach for Reproductive Equity, a measurement framework for human-AI interaction that adds capability outcomes as a unit of evaluation above task performance. CARE functions in two parts. The Normative Design Lens identifies the resources, conversion factors, capabilities, and functionings a system should support. The Evaluation lens assesses how design features, interaction patterns, and social conditions shape capability outcomes, tradeoffs, and lived experiences in use. We apply CARE to SRH-specific chatbots, general-purpose LLMs, and search engine features in a study with 12 participants, demonstrating that it surfaces capability outcomes standard metrics aggregate away. The same design features expanded capabilities for some users while constraining them for others: source-level organization, response format, tone, and SRH-specific features all shaped which capabilities expanded for which users and in which direction. Participants' professional backgrounds, gender identities, and prior AI familiarity further shaped these effects, producing capability outcomes that usability and accuracy metrics, aggregated across users, would not surface. These findings demonstrate capability outcomes as a measurable unit for human-AI interaction evaluation, extending existing metrics with a capability layer above task performance.
\end{abstract}

 
\keywords{Human-AI interaction, Design and Measurement of AI System, Capability approach \vspace{3em}\\}

\maketitle

\section{Introduction}
\begin{figure*}[h]
  \centering
  \includegraphics[width=0.8\textwidth]{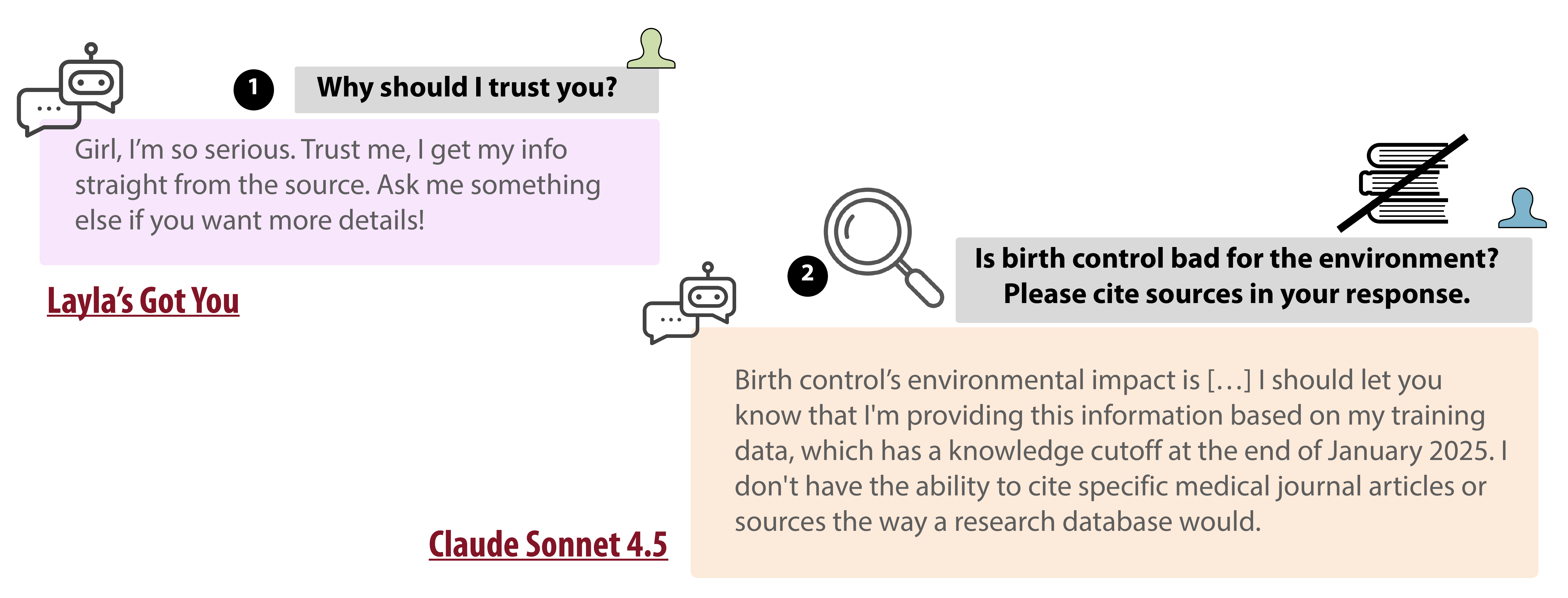}
  \caption{\textbf{Source transparency as a capability condition.} 
  Layla's Got You (left) presents information without source attribution, building implicit trust without verification support. Anthropic Claude Sonnet 4.5 (right) presents source limitations as a system constraint, discouraging source-seeking behavior. Both design decisions function as conversion factors: they determine whether the tool's information becomes a capability for the user or remains a resource they cannot act on, the distinction CARE's Evaluation Lens measures.}
  \label{fig:source_transparency}
\end{figure*}

Algorithmic systems now mediate sexual and reproductive health (SRH) information seeking for many people navigating stigma, limited access to care, and uneven access to medically accurate guidance \cite{Tuli2019, Sorcar2017, Jain2015, Moin2025, Tuli2023}. Globally, 2 in 3 girls enter puberty without the knowledge they need \cite{unesco_cse_puberty}, only 41\% of countries provide in-school SRH education \cite{anderson2025biomedical}, and 71\% of individuals aged 15--24 report the internet as their primary source of SRH information \cite{unesco_cse_puberty, martinez2025105}. SRH chatbots 
\cite{bonnevie2021layla, Rahman2021, woo2020development, workPlannedParenthood} respond to this demand with anonymous, around-the-clock support for stigmatized topics, shaping what users can verify, understand, and act on. These design choices operate at a level that scale and access statistics alone do not capture (Figure \ref{fig:source_transparency}).

Standard HCI and AI evaluation centers usability, accuracy, and interaction quality\cite{nielsen2005ten, amershi2019guidelines}, measures designed to assess task performance and interaction quality at the system level. A concise answer may support action for one user and suppress source-checking for another, while a supportive interface may reduce stigma for some and exclude others through assumptions about gender, sexuality, or access to care. HCI has applied the capability approach to reframe design questions for care workers and smart systems \cite{khullar2025nurturing, heath2019relations, bondi2021envisioning}, and scholars have argued that evaluating generative AI requires methods that capture socially situated outcomes \cite{wallach2025position}. Neither line of work has operationalized these insights into design and evaluation lenses for SRH-AI interaction specifically.

We introduce CARE, the Capability Approach for Reproductive Equity, to address this gap. CARE brings Sen's capability approach \cite{Sen1999, sen2004health} into SRH technology design and evaluation as an organizing philosophy. Sen defines development as the expansion of real freedoms, a standard that shifts the measure of success from resource provision to the opportunities resources actually create. Under this framing, a chatbot succeeds when health literacy, healthcare access, stigma, and legal constraints allow users to convert its information into meaningful action. Those conversion conditions determine whose freedoms a tool actually expands. Current SRH chatbot design reflects resource-centered thinking: developers build tools around features and information delivery, treating capability support as a secondary concern rather than the design starting point. Mainstream SRH chatbots often adopt female personas \cite{mills2024chatbots}. These designs reinforce norms that frame reproductive health as solely a women's concern and cut off users of other genders. Acting on a tool's guidance demands money, literacy, trust, and social safety that many users lack. Reproductive equity depends on these systems expanding people's real freedom to make informed reproductive decisions.

Measuring whether SRH tools expand real freedoms requires extending existing evaluation methods beyond task completion and error rates. Usability, accuracy, and interaction quality measures remain essential; extending them to the conditions through which users convert access into capabilities and capabilities into meaningful outcomes makes them more complete. As AI systems take on a larger role in SRH information seeking, evaluation methods need to account for that conversion layer.

CARE includes two complementary lenses that together close the loop between design intent and lived outcome, a connection that prior capability-based health technology work has theorized but not operationalized at the level of specific design features and interaction patterns \cite{khullar2025nurturing, heath2019relations, bondi2021envisioning}. The Normative Design Lens starts from reproductive wellbeing and works backward to identify what SRH technologies should support. The Evaluation Lens starts from resources and works forward to assess how design features, interaction patterns, and social conditions shape capability outcomes, tradeoffs, and lived experiences in use.

We make three contributions. First, we introduce CARE as a measurement framework for human-AI interaction (HAI) that adds capability outcomes as a unit of evaluation above task performance, instantiated here for SRH contexts. Second, we apply CARE to SRH-specific non-LLM chatbots, general-purpose LLMs, and search engine features in a study with 12 participants. The same design features expanded capabilities for some users while constraining them for others: source-level organization, response format, tone, and SRH-specific features all shaped which capabilities expanded for which users and in which direction. Participants' professional backgrounds, gender identities, and prior AI familiarity further shaped these effects, producing capability outcomes that usability and accuracy metrics, aggregated across users, would not surface. Third, we derive evaluation and design recommendations, and policy implications from these findings, including measurement criteria for capability outcomes, evaluation strategies that center affected populations, and regulatory requirements for AI health guidance.

\begin{figure*}[ht!]
  \centering
  \includegraphics[width=0.95\textwidth]{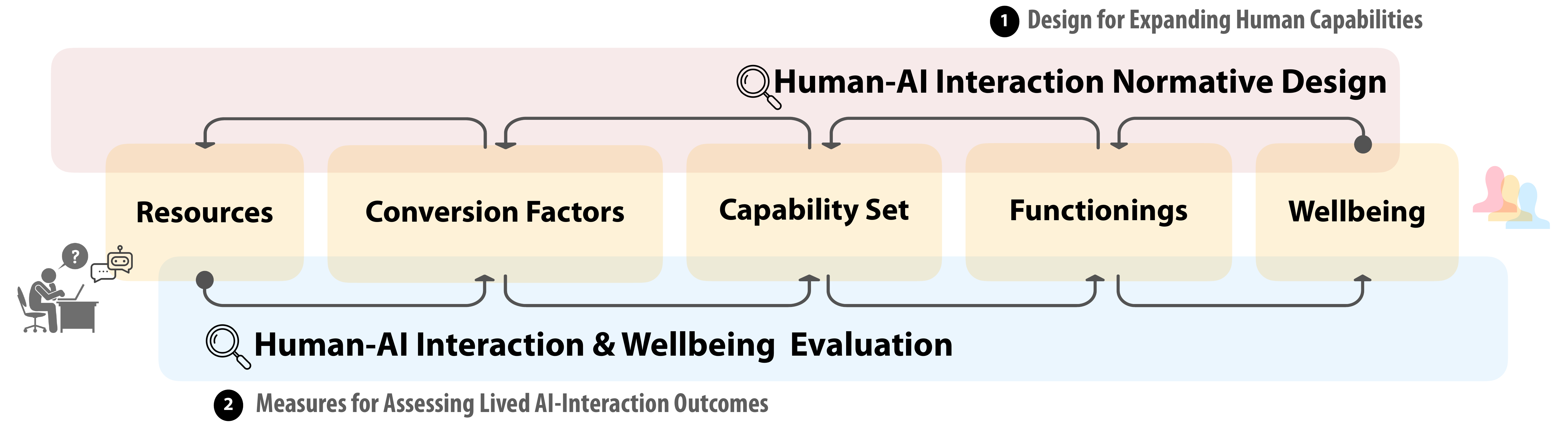}
\caption{\textbf{CARE Normative Design and Evaluation Lenses.} The Normative Design Lens starts with wellbeing, defined here as reproductive freedom, and identifies the functionings, capabilities, conversion factors, and resources required to support that goal by working backward through the capability chain. The Evaluation Lens starts with resources and works forward to assess how design features, interaction patterns, and sociotechnical conversion factors shape capabilities, how capabilities are expressed in functionings, and how these processes affect reproductive wellbeing. This forward lens reveals human-AI interaction tradeoffs and differences in lived outcomes across users (see Table \ref{tab:nussbaum_beings_doings} for the capability set and example functionings).}
  \label{fig:Capability Approach Diagram}
\end{figure*}

\section{Related Work} 
CARE sits at the intersection of three bodies of work: capability-based approaches to health equity in sociotechnical systems, information seeking behavior and trust in SRH chatbots, and evaluation frameworks for SRH technology.  

\subsection{The Capability Approach in Sociotechnical Systems} The WHO defines health equity as achieved when "everyone can attain their full potential for health and well-being" \cite{WHOHeathEquity2025}. Sen extends this by characterizing health equity as a multidimensional concept integrating health outcomes, delivery processes, and structural conditions \cite{Sen1999, sen2004health}. The capability approach evaluates wellbeing through four linked concepts: resources (what users can access), conversion factors (personal, social, and structural conditions that mediate access), capabilities (real freedoms to achieve desired outcomes), and functionings (outcomes actually achieved) \cite{sen1993capability, sen1999commodities, robeyns2017wellbeing}. A person may have access to an SRH chatbot (resource), but stigma or limited health literacy (conversion factors) may prevent them from making informed reproductive decisions (capabilities). Those barriers keep opportunities from producing real health outcomes (functionings). Martha Nussbaum extended this by specifying Ten Central Human Capabilities as a threshold for a dignified life \cite{nussbaum2000women}, criteria that have anchored capability-based health research \cite{venkatapuram2013health, mitchell2017applications} and HCI applications of the approach.

Within HCI and AI, scholars have applied the capability approach to reframe the design question from "what does this technology provide?" to "what freedoms does it create?" This shift has guided work on gaps between AI design and user aspirations in care workers \cite{khullar2025nurturing}, capabilities enabled by smart systems \cite{heath2019relations}, and participatory design for marginalized groups \cite{bondi2021envisioning}. The HCI community has also studied how structural determinants such as race and age shape health equity \cite{parker2025participatory, iloanusi2024ai, guan2021taking, jo2024understanding}. These approaches have advanced understanding of how individual determinants shape health equity; CARE extends this by connecting those determinants to upstream structural conditions through the capability approach. Prior applications establish which freedoms matter and demonstrate the capability approach as a productive lens for technology design; CARE takes the next step by specifying which capabilities a given system should support and measuring whether it produces them for specific users. The Normative Design Lens specifies the capability set SRH technologies should support, and the Evaluation Lens measures whether design features and social conditions produce those outcomes in practice.

\subsection{Information Seeking, Trust, and Stigma in SRH Contexts} 
Conversion factors such as stigma, health literacy, and social safety appear concretely in research on information seeking and trust in SRH contexts \cite{Tuli2019, Sorcar2017, Tuli2023, dewan2024teen, kaleva2026privacy}. Information seeking is an iterative process shaped by information load, source credibility judgments, and the effort required to evaluate diverse sources \cite{kaiser2025new, yang2025search}. Users trade depth for efficiency, relying on familiar tools or the first credible result \cite{capra2023does}. In SRH contexts these general challenges compound: emotion, stigma, and privacy concerns all shape both what people search for and whether they seek information at all\cite{xiao2023powering, dewan2024teen}.

Cultural taboos create additional barriers. Stigma surrounding menstruation leads teachers to skip related content and students to develop secret vocabularies around the topic \cite{Tuli2019}. Some regions ban HIV education entirely \cite{Sorcar2017}. Men face systematic exclusion from SRH education at home and at school, producing information gaps that extend across genders \cite{Tuli2023}. These gaps push individuals toward anonymous digital sources, where exposure to misinformation and delayed care increases \cite{dewan2024teen}. AI chatbots have responded by offering 24/7 support, anonymity, and conversational interfaces that reduce the social cost of asking stigmatized questions \cite{kaleva2026privacy, xiao2023powering}.

Users develop trust in these systems through surface cues such as perceived accuracy, transparency, and conversational tone \cite{kaleva2026privacy, morosoli2025transparency}. Trust forms across source, content, and system levels and adjusts over time \cite{kaleva2026privacy}. Transparency features such as AI disclosure labels and factuality indicators reduce over-reliance in some cases and increase skepticism in others \cite{morosoli2025transparency, do2025hide, do2025highlight}. In stigmatized SRH contexts, fears of surveillance or criminalization constrain engagement even when users find a chatbot credible \cite{kaleva2026privacy}. This body of work has advanced trust calibration and interface design at the interaction level. CARE treats stigma, privacy fears, and health literacy as conversion factors: structural conditions that shape whether information access produces reproductive capabilities for a given user.

\subsection{SRH Technology Design and Evaluation} 
SRH chatbot developers have worked to address these trust and access barriers directly, producing a body of tools and evaluation approaches that CARE builds on and extends. Digital platforms such as Menstrupedia and TeachAIDS provide menstrual health and HIV education \cite{Tuli2018, Sorcar2017}. Visual and interactive media have improved understanding of topics such as cervical cancer and female anatomy \cite{Moin2025, Tran2018}. AI chatbots have emerged as tools to deliver accessible, culturally sensitive SRH information to adolescents and underserved populations \cite{Rahman2021, Deva2025}. Usability heuristics \cite{nielsen2005ten} and HAI guidelines \cite{amershi2019guidelines} have advanced understanding of accessibility and interaction quality across these tools.

Emerging work argues that evaluating generative AI requires new measurement methods that capture socially situated outcomes \cite{wallach2025position, Sharma2026}. Reducing sensitive health topics to technical problems produces specific distortions: the false promise of objectivity, the presumption of biomarkers as universal, and assumptions about unidirectional disease transmission \cite{ma2023you, singh2022automating}. These contributions establish a foundation for assessing SRH tools along a broader dimensions of usability, accuracy, and interaction quality.

Evaluating whether those tools expand users' reproductive freedoms requires a principled account of which capability outcomes they should produce, a design question CARE addresses directly. Design decisions about tone, source structure, persona, and response format each carry capability consequences that task-level metrics were designed to measure at the interaction level. CARE adds the structural layer above them by introducing capability outcomes as a unit of measurement, grounding both design and evaluation in Sen's capability approach. This is critical to connect specific design features and interaction patterns to the conditions that determine whose reproductive freedoms a tool expands.


\begin{table*}[!ht]
\centering
\renewcommand{\arraystretch}{1.5}

\caption{Normative mapping of Nussbaum's Ten Human Capabilities to SRH-specific beings and doings. *We select all but two capabilities (Other Species and Play) for our evaluation. }

\begin{tabular}{
p{0.20\textwidth}
p{0.35\textwidth}
p{0.35\textwidth}
}

\hline
\textbf{Nussbaum Capability} &
\textbf{SRH-specific Beings (States of Being)} &
\textbf{SRH-specific Doings (Actions)} \\ \hline

Life &
Being able to live a normal-length life without preventable death from pregnancy or sexually transmitted infections (STIs). &
Preventing unintended pregnancy; accessing safe abortion services and STI treatment. \\ \hline

Bodily Health &
Being sexually and reproductively healthy, well-nourished, and physically safe. &
Using contraception effectively; obtaining STI testing and treatment. \\ \hline

Bodily Integrity &
Being in control of one’s body and reproductive choices. &
Choosing if and when to have sex or children; accessing abortion or contraception. \\ \hline

Senses, Imagination, and Thought &
Being informed through accurate, non-stigmatizing sexual and reproductive health knowledge. &
Learning about SRH; understanding risks, methods, and available options. \\ \hline

Emotions &
Being free from fear, shame, or anxiety related to sex and reproduction. &
Seeking care or information without embarrassment or emotional distress. \\ \hline

Practical Reason &
Being able to form, reflect on, and pursue one’s own reproductive goals. &
Making informed decisions about contraception, sexual activity, and pregnancy. \\ \hline

Affiliation &
Being socially included and respected regardless of gender, age, or identity. &
Discussing SRH with partners or providers; avoiding stigma and discrimination; seeking care without social sanctions or judgment. 
\\ \hline

Other Species* &
Less central in this context, but relevant to environmental considerations of health. &
Considering environmental impacts of health-related choices where applicable. \\ \hline

Play* &
Being able to experience pleasure, intimacy, and joy safely. &
Engaging in consensual sexual relationships without fear of harm. \\ \hline

Control over One’s Environment &
Being able to access sexual and reproductive health services and legal rights. &
Obtaining affordable birth control; using insurance coverage or telehealth services. \\ \hline

\end{tabular}
\label{tab:nussbaum_beings_doings}
\end{table*} 

\section{CARE: A Capability-Based Measurement Framework for Reproductive Equity in Human-AI Interaction} 
\label{sec:care} 
CARE introduces capability outcomes as a unit of measurement for HAI, connecting interaction design decisions to the structural conditions that determine whose freedoms a system expands. It produces actionable design and evaluation guidance by traversing the capability chain in both directions: the Normative Design Lens works backward from reproductive wellbeing to specify what tools should support, and the Evaluation Lens works forward from resources to measure whether they do. CARE operationalizes Sen's distinction between capabilities and functionings through Robeyns' core (A) and specification (B) modules, treating conversion factors as the mechanism linking resources to real freedoms (see Appendix Capability Approach Modules for details) \cite{robeyns2017wellbeing}. 

CARE is tailored to SRH in three ways. First, it specifies a capability subset from Nussbaum's Ten Central Human Capabilities \cite{nussbaum2000women} relevant to reproductive contexts (see Table \ref{tab:nussbaum_beings_doings}). Second, it foregrounds conversion factors specific to SRH: reproductive stigma, gendered design assumptions, and legal constraints on reproductive autonomy. Third, it defines wellbeing in terms of reproductive freedom rather than general health outcomes. These specifications anchor CARE in SRH. The underlying structure adapts to other sensitive domains where conversion factors shape capability outcomes. 

The framework operates through two lenses, each structured around resources, conversion factors, capabilities, and functionings. \textit{The Normative Design Lens} specifies how SRH tools should be designed by following the capability approach \textbf{backwards}: (1) establish wellbeing goals grounded in reproductive freedom, (2) specify desired functionings as target ``beings'' and ``doings'', (3) identify capabilities that enable those functionings, (4) determine conversion factors that shape users' ability to achieve those capabilities, and (5) design resources intended to enable those capabilities. \textit{The Evaluation Lens} follows the capability approach \textbf{forward} to assess existing technologies: (1) analyze the technology as a resource, (2) examine how conversion factors shape users' ability to convert resources into capabilities, (3) evaluate the technology's impact on the capability set, (4) assess the functionings users actually achieve, and (5) assess the resulting impact on SRH wellbeing. 

We first apply the Normative Design Lens to establish the capability baseline against which existing tools are evaluated. The Evaluation Lens is applied in the study reported in the section CARE: Evaluation Lens.

\section{CARE: Normative Design Lens} 
\label{sec:normative} 

\subsection{Capability Baseline for SRH Tools}
Applying the Normative Design Lens to SRH produces the baseline against which we evaluate existing tools. Working backward from reproductive freedom (Step 1: establishing well-being goals), the lens identifies desired functionings such as making informed contraceptive decisions, understanding fertility risks, and seeking care without shame (Step 2). These functionings require capabilities.
The framework allows us to use an existing capability set or create a new one with community input. We chose to use Nussbaum's Ten Central Human Capabilities \cite{nussbaum2000women} because they are well aligned with reproductive contexts. We subset the capabilities most relevant to SRH, including Practical Reason, Bodily Health, Bodily Integrity, and Control over One's Environment (Step 3). Conversion factors determine whether individuals can actually realize them. Low health literacy, reproductive stigma, and limited trust in AI-generated information each shape users' ability to convert those resources into reproductive capabilities (Step 4). Technologies designed to enable these capabilities should therefore provide literacy-calibrated responses with verifiable sources and non-stigmatizing language (Step 5).

To understand how individuals realize capabilities in practice, we used the 369 frequently-asked, expert-validated-and-answered SRH questions from \textit{bedsider.org}. In order to subset the list of questions to use for evaluation, we mapped the 369 questions to a subset of Nussbaum's Ten Central Human Capabilities and grouped the questions into 11 question categories based on their capability mapping. Table \ref{tab:questionsfunctioningcapabilities} presents the normative mapping of the 11 question categories to their associated functionings and capabilities. This mapping operationalizes reproductive freedom as a measurable outcome, establishing the capability baseline the study applies to SRH-specific chatbots, general-purpose LLMs, and search engine features.

\begin{table*}[!h]
\renewcommand{\arraystretch}{1.2}
\centering
    \caption{Normative mapping of question categories to their associated functionings and capabilities through the Normative Design Lens (from wellbeing to resources). The functionings shown are examples and not an exhaustive list.}
\begin{tabular}{p{4cm} p{6cm} p{6cm}}
\toprule
         \textbf{Question Category} &\textbf{Functionings} & \textbf{Capabilities} \\
\midrule
         Birth Control Methods \& Use& Correctly using a chosen contraceptive method and managing one's fertility & Bodily Integrity, Practical Reason, Control over One's Environment (Material)\\
\midrule         
         Effectiveness and Pregnancy Risk& Avoiding unintended pregnancy and understanding fertility and pregnancy risk& Life; Bodily Health; Practical Reason\\
\midrule         
         Side Effects, Safety, and Health Risks& Maintaining physical and mental well-being and managing side effects safely & Bodily Health, Life, Senses, Imagination, and Thought\\
\midrule
         Sexual Health, STIs, and Protection& Preventing STIs, getting tested and treated, and engaging in safer sexual practices & Bodily Health, Bodily Integrity, Life\\
\midrule
         Emergency Contraception and Abortion &  Preventing pregnancy after sex, ending and unwanted pregnancy safely  & Bodily integrity, Life, Practical Reason, Control over One's Environment (Political)\\
\midrule
         Access, Cost, and Insurance& Obtaining affordable reproductive healthcare and using services without financial hardship  & Control over One's Environment (Material), Bodily Health, Affiliation (Non-discrimination)\\
\midrule
         Over-The-Counter (OTC) and Opill& Independently obtaining birth control and timely pregnancy prevention & Control over One's environment, Practical Reason, Bodily Integrity\\
\midrule
         Relationships, Sex, and Communication&  Communicating sexual boundaries and experiencing consensual intimacy  & Affiliation, Bodily Integrity, Emotions\\
\midrule
         Identity, Inclusivity, and Special Populations& Accessing appropriate care regardless of identity or status and being respected in health systems  & Affiliation, Bodily Integrity, Control over One's Environment\\
\midrule
         Misinformation, Myths, and Stigma&  Acting on accurate sexual and reproductive health information and seeking care without shame or fear  & Senses, Imagination, and Thought, Practical Reason, Affiliation\\
\midrule
         Reproductive Anatomy, Terms, and Education & Understanding one's body and health and interpreting medical information & Senses, Imagination, and Thought, Practical Reason\\
\bottomrule
    \end{tabular}    \label{tab:questionsfunctioningcapabilities}
\end{table*}

\section{CARE: Evaluation Lens} 
\label{sec:study} 

To validate CARE as a measurement framework, we evaluated three SRH tools against the capability baseline established in the previous section, measuring capability outcomes across tools, users, and conversion factors. This study was determined exempt from IRB review by our institution's Institutional Review Board.

\subsection{Participants} 
We recruited twelve participants in-person and via email. All participants provided recording consent prior to participation. Participants first completed a brief questionnaire covering their experience with SRH education, information-seeking habits, experience with AI chatbots, and demographic information. Participants ranged in age from 18 to 62. Nine self-reported their gender as female, two as male, and one as non-binary (see Table \ref{tab:participants}).

 \begin{table}[h!]
    \centering
    \small
    \caption{Participant details (n = 12). M = male, F = female, and NB = non-binary. Literacy scores were self-rated on a 5-point Likert scale with 1 being the most uncomfortable and 5 being the most comfortable.}
    \begin{tabular}{>{\centering\arraybackslash}p{0.6cm}>{\centering\arraybackslash}p{0.4cm}>{\centering\arraybackslash}p{1.15cm}>{\centering\arraybackslash}p{1.2cm}>{\centering\arraybackslash}p{1.2cm}>{\centering\arraybackslash}p{1.2cm}}\toprule
         \textbf{Alias}&  \textbf{Age}&  \textbf{Gender Identity}&   \textbf{SRH Literacy}&\textbf{AI Literacy}&\textbf{English Literacy}\\\midrule
         P1& 21 & M &   3&4&4\\\midrule
         P2& 39 & F &   4&4&5\\\midrule
         P3& 21 & M &   3&4&4\\\midrule
         P4& 22 & F &   4&5&5\\\midrule
         P5& 28 & F &   2&3&5\\\midrule
         P6& 25 & F &   5&5&4\\\midrule
         P7& 22& F &   4&3&4\\\midrule
         P8& 22& F &   4&3&5\\\midrule
         P9& 20& F &   4&4&5\\\midrule
         P10&  62& F &   5&2&5\\\midrule
         P11& 25 & NB &   3&1&5\\\midrule
         P12&  18& F &   5&3&4\\ \bottomrule
    \end{tabular}
    \label{tab:participants}
\end{table}

\subsection{Tools}
After an initial exploration of multiple non-LLM SRH-specific chatbots and general-purpose LLMs, we selected three tools representing distinct categories of SRH information systems: Layla's Got You \cite{bonnevie2021layla}, a non-LLM SRH-specific chatbot; OpenAI ChatGPT-4o mini \cite{openai_chatgpt_4omini_2026}, a general-purpose LLM; and Google Search \cite{google_search_engine2026}, a search engine. We chose OpenAI ChatGPT-4o mini over Anthropic Claude Sonnet 4.5 \cite{anthropic2024claude} due to its higher popularity. These tools span the range of digital systems people currently use for SRH information seeking.

\subsection{Question Selection} 

To select questions for our evaluation, we started the 11 question categories developed using the Normative Design Lens
(see Table \ref{tab:questionsfunctioningcapabilities}). We selected one question from each of four categories, choosing the smallest set of categories that covered all unique capabilities: (1) Birth Control Methods and Use, (2) Effectiveness and Pregnancy Risk, (3) Side Effects, Safety, and Health Risks, and (4) Relationships, Sex, and Communication. 

\subsection{Procedure} 
We conducted each session individually over Zoom. Each participant researched one question across all three tools, a design that prioritized depth over breadth. Participants signed out of ChatGPT-4o mini and Google Search and remained in incognito mode throughout to reduce variability that may be present based on previous use of the tools.  

We assigned participants to four SRH questions across four topic areas, with three participants per question. In the category of Birth Control Methods and Use, participants 1, 5, and 9 were asked, \textit{``How long does the IUD last?''} In Effectiveness and Pregnancy Risk, participants 2, 6, and 10 were asked, \textit{``Can I get pregnant on my period?''} In Side Effects, Safety, and Health Risks, participants 3, 7, and 11 were asked, \textit{``Does birth control cause depression?''} In Relationships, Sex, and Communication, participants 4, 8, and 12 were asked, \textit{``My partner can feel my IUD strings—what do I do?''}

Participants received links to all three tools and used them in any order. For each tool, participants had up to 10 minutes to interact, then gave a brief summary of what they learned. After each tool interaction, participants assessed it for the associated capabilities. During each tool interaction, participants shared their screen with the researcher, who took notes on search strategies and interaction patterns. After completing all three tools, participants took part in a brief semi-structured interview covering response clarity, likes and dislikes about each tool's responses, and perceptions of how each tool functioned. Each session lasted approximately one hour and was transcribed using Zoom. Two researchers independently coded the transcripts and tagged with Tagguette \cite{Rampin2021}.

\subsection{Analysis} 
Three researchers analyzed the session transcripts in Taguette \cite{Rampin2021} using an inductive coding process. Codes captured interactions patterns which were then grouped and mapped back to CARE's capability set. The team then used reflexive thematic analysis to discuss interpretations and iteratively developed themes \cite{braun2019reflecting}. This process produced five themes: Agency and Control, Trust and Credibility, Information Processing and Learning, Identity and Ideology, and Usability and Interaction.

\section{Results: Evaluation Lens Findings} \label{sec:results} Applying CARE's Evaluation Lens to the three tools revealed how interaction design patterns produce capability consequences, how conversion factors determine which consequences materialize for which users, and how the same design choice expanded capabilities for some participants while constraining them for others. Against the normative baseline established in the section CARE: Normative Design Lens, all three tools produced partial capability support depending on who was using them. 

\subsection{Interaction Patterns and Their Capability Consequences} Each interaction pattern traces to a specific design decision and produces a specific capability consequence. CARE's Evaluation Lens makes this chain visible: the same behavior that reads as a usability observation at the interaction level reveals a capability outcome at the structural level. 

\subsubsection{Source-Level Organization Expands Practical Reason and Control over One's Environment} Presenting information organized by source makes source origin visible before content, enabling users to selectively engage with credible sources. All participants checked at least two sources during Google searches. P12 described corroboration as the basis for perceived accuracy: \textit{``Because I feel like the more similar responses there is, that means it's like more accurate and supported.''} P7 used domain extensions to filter credibility: \textit{``I know Harvard Health, it says .edu, so I would open that first.''} P5 described the resulting agency: \textit{``I had the most control over what information to take in, because I can actually click into each of those links.''} Source-level organization in Google expanded Practical Reason and Control over One's Environment by giving users the evaluative infrastructure to assess information before accepting it. Where that infrastructure was absent, as with Layla's Got You, trust dropped. P5 stated: \textit{``On Layla, I would not trust it as much because I don't have anything to corroborate the answers from there with.''} 

\subsubsection{Source-Level Organization Constrains Senses, Imagination, and Thought Through Information Load} The same source-level structure requires users to synthesize across self-contained chunks, producing skimming that prevents deep processing. For example, participants read only abstracts when clicking into academic studies found on Google. P5 described the effort cost of reading a source: \textit{``there's a whole 5,000 words of information on each webpage, so that is extra effort cost.''} P6 found the volume unmanageable: \textit{``it just gives too many statements at the one time... too much effort to understand it.''} Interface complexity in Layla's Got You produced the same pattern independent of information volume. P4 described the website as having \textit{``a lot of tabs, a lot happening''} and asked the chatbot after failing to find information through the interface. Skimming constrained Senses, Imagination, and Thought by keeping users at the surface of information rather than building the understanding needed to act on it. 

\subsubsection{Expert-Credibility Filtering Reflects Conversion Factors Shaping Affiliation} Most participants avoided forum sites during Google searches, reading non-expert sources as threats to information quality. P4 stated \textit{``Healthline feels like the better source than Reddit''} and P10 stated \textit{``I would not click on any of these Reddits.''} P8 was the only participant who engaged with Reddit: \textit{``I like seeing what other people, common folk, are experiencing in a sexual and reproductive health situation.''} The divergence reveals a conversion factor at work: prior SRH experience shaped whether community knowledge expanded or constrained Affiliation. Designs that surface only expert sources by default encode one conversion profile as the standard user. 

\subsubsection{Keyword-Matching Architectures Constrain Practical Reason, Senses Imagination and Thought, and Emotion} Architectures that return fixed responses triggered by keywords constrain three capabilities through a single design decision. Using Layla's Got You, P2 described losing trust immediately: \textit{``it just seemed to be in a cycle of the same answer.''} P3 concluded their research after recognizing the pattern: \textit{``I think my research does kind of conclude there now that I have a familiarity that Layla tends to respond with identical responses.''} P10 summarized: \textit{``anybody in their right mind would not keep typing in this.''} Frustration constrained Emotion, the inability to obtain relevant information constrained Practical Reason, and the absence of new content constrained Senses, Imagination, and Thought — all three tracing to Layla's fixed response matching. 

\subsubsection{Confident Tone with Absent Citations Produces Passive Source Acceptance} Response designs that lead with a confident tone and present conclusions without visible citations build implicit trust that suppresses source verification. Using ChatGPT, P3 identified the mechanism: \textit{``It used language that convinced me that GPT knows what it's talking about. Whether it used real sources or not, it didn't really matter. Seeing such extensive text about concepts I was learning for the first time, I felt I could trust this GPT model without asking it for sources.''} P7 acknowledged the same dynamic: \textit{``I know I can't really trust it that much, but because it's given to me so confidently, I do trust it.''} Passive acceptance constrained Practical Reason by bypassing source verification and constrained Control over One's Environment by removing users' ability to evaluate the basis of what they received. When P11 prompted ChatGPT for \textit{``birth control pill causing depression statistics" } and received citations, source evaluation resumed: \textit{``[ChatGPT] did have links to, for example, the Harvard study.''} 

\subsubsection{Default Clinical Redirection Constrains Practical Reason at the Point of Need} Designs that redirect users to consult a clinician before providing actionable information constrain Practical Reason by closing the decision-making loop before the user has what they need to act. P4 found Layla's redirection credible: \textit{``I trusted them because they were mainly just saying talk to a physician.''} P12 described the failure for urgent situations: \textit{``if I'm in the middle of sex... they tell me to go see my doctor... I'm just going to ignore it and go to a different site.''} This design encodes the assumption that users have immediate access to clinical care, a conversion factor that does not hold for all users. 

\subsubsection{Query Refinement Reflects Effort to Recover Constrained Capabilities} When interaction designs failed to answer specific queries, participants rephrased questions to recover the capability the design had not produced. P1 described using Layla's Got You as \textit{``a game of cat and mouse about trying to find the right question to get the right answer out.''} P11 tried keyword specification: \textit{``I'm just going to type in depression, just to see if it triggers anything.''} Query refinement represents users working around a constrained resource rather than converting it into a capability, effort the normative baseline assigns to the design, not the user. 

\subsection{Conversion Factors: How Individual Conditions Shaped Capability Outcomes} Professional background, gender identity, and prior AI familiarity functioned as conversion factors: they determined whether the same design feature produced a capability for one participant and left it inert for another. Capability outcomes cannot be evaluated against an average user. Evaluation requires specifying the conversion factors present in the target population. 

\subsubsection{Professional Background} Healthcare background functioned as a validation framework that amplified capability expansion from source-transparent designs. P2 used prior clinical knowledge to assess response alignment: \textit{``because I have some background in healthcare, I do kind of have a pre-existing opinion on this. So as I'm reading this, I'm thinking, okay, this does align with what I expected.''} P10 applied a clinical frame to urgency and directed attention to whose needs the designs addressed: \textit{``none of them considered that there might be the implicit bias that they didn't consider same sex.''} Professional background expanded Practical Reason by supplying a validation infrastructure most users lacked, a conversion factor designs evaluated against a general user profile would not surface.

\begin{table*}[ht!]
    \centering
    \caption{Selected dimensions where CARE adds a capability 
    measurement layer above standard HAI evaluation (e.g., Microsoft Human-AI Interaction Guidelines \cite{amershi2019guidelines}), grounded 
    in study findings. Rows show cases where interaction-level 
    evaluation alone would miss capability outcomes that varied 
    across users.}
    \begin{tabular}{p{4cm}p{4cm}p{8.5cm}}
    \toprule
    \multicolumn{1}{c}{\textbf{HAI Guideline} 
   } &
    \multicolumn{1}{c}{\textbf{Interaction-Level Measure}} &
    \multicolumn{1}{c}{\textbf{Capability-Level Measure}} \\
    \midrule
    G6: Mitigate social biases & 
    The system avoids overt bias and exclusionary language & 
    Users with marginalized identities feel recognized and safe, 
    and implicit design assumptions are surfaced. \textit{``Support 
    for people like me''} (P11, Google); \textit{``all of this 
    discussion of sex is in the male female paradigm''} (P10, 
    Google) \\
    \midrule
    G11: Make clear why the system did what it did & 
    The system shows its reasoning or sources & 
    Users can verify, compare, and inspect information sources, 
    retaining evaluative control over what they receive. 
    \textit{``If you have to explicitly ask what sources did you 
    use? Then I think it's already suspicious.''} (P3, ChatGPT) \\
    \midrule
    G4, G10: Scope services when in doubt; G16: Convey the 
    consequences of user actions & 
    The response gives next-step guidance & 
    Users can act meaningfully after the interaction with 
    sufficient understanding to make an informed reproductive 
    decision. \textit{``[Google] allowed me to think about how 
    to act, but never taught me how to act.''} (P3, Google) \\
    \midrule
    G8: Support efficient dismissal; G17: Provide global 
    controls & 
    Users can enter, leave, navigate, and steer the interaction & 
    Users retain substantive autonomy to choose, compare, inspect, 
    and reject information sources. \textit{``It feels like it 
    gives me a sense of agency''} (P1, Google); \textit{``you 
    sacrifice the ability to choose the sources''} (P3, ChatGPT) \\
    \midrule
    Substantive human outcome \newline
    \textit{(no equivalent HAI guideline)} & 
    Task completion and interaction quality & 
    What the user can now know, decide, feel, and do: practical 
    reason, bodily integrity, emotion, affiliation, and 
    informational control. \textit{``More empowered\ldots more 
    sure of my decision.''} (P2, ChatGPT) \\
    \bottomrule
    \end{tabular}
    \label{tab:new_eval}
\end{table*}

\subsubsection{Gender Identity and Intersectionality} Gender identity shaped whether information felt personally applicable, affecting Bodily Integrity and depth of engagement. P3, who identifies as male, stated: \textit{``I definitely have to approach it from the third person. I have no way of using, say, like an IUD.''} This distancing constrained Bodily Integrity because the information carried no personal applicability. P11, who identifies as non-binary, evaluated Google sources through perceived institutional care for their identity: \textit{``Since I'm non-binary, there's a lot of transphobia in the medical field. So I'm much more inclined to go to sites like Planned Parenthood that have explicitly stated a support for people like me.''} Source-transparent designs expanded Affiliation and Control over One's Environment for P11 through identity-driven source selection, a conversion the same design did not produce for participants without that evaluative frame. 

\subsubsection{Prior AI Familiarity} Prior AI use built brand trust that determined whether confident response tone expanded or constrained Practical Reason. P6 and P7, who had used ChatGPT for health queries, extended that trust into the study. P7 stated: \textit{``I know I can't really trust it that much, but because it's given to me so confidently, I do trust it.''} Brand trust suppressed source-seeking and constrained Practical Reason. P11's skepticism produced the opposite outcome: \textit{``it did cite sources, but I would want to check out the sources myself.''} The same ChatGPT response, two opposite capability outcomes, determined by a conversion factor the design had no mechanism to detect. 

\subsection{Cross-tool Tradeoffs and Functionings} The most consistent finding was that the same design feature expanded capabilities for some participants while constraining them for others. CARE makes this visible by tracing outcomes at the individual level. Standard usability metrics aggregated across users would not surface these opposing-direction effects. \textbf{Practical Reason} produced the sharpest tradeoffs. Google's source-transparent design expanded it for users with source evaluation capacity and introduced effort costs that constrained it for users who needed fast actionable answers. P3 captured this: \textit{``Google is more informative, which means that it allowed me to think about how to act, but never taught me how to act.''} ChatGPT's BLUF structure expanded Practical Reason through actionable responses while absent citations constrained it by suppressing verification. Layla's default clinical redirection constrained Practical Reason by withholding actionable information at the point of need. \textbf{Affiliation and Emotion} revealed the clearest within-tool tradeoff. Layla's empowerment-oriented design expanded Affiliation and Emotion for participants who read its aesthetic as inclusive. P8 explained: \textit{``it felt like this is made by a team of women. This is made with empowerment in mind.''} Layla's keyword-matching architecture in the same tool then constrained Emotion by producing frustration when repeated responses failed to meet the expectations raised by the welcoming interface. Two independent design systems in the same tool produced opposing capability effects, a structural tension CARE surfaces that usability ratings would average away. \textbf{Control over One's Environment} divided across designs by source visibility. Google's source-transparent design expanded it for users with source evaluation capacity. ChatGPT's single-response format constrained it by closing interaction after the first answer. P3 identified the tradeoff: \textit{``ChatGPT, while sacrificing that ability to choose sources, provides a summary in layperson terms and easily interpretable and slightly actionable.''} A \textbf{Functionings} assessment completed the Evaluation Lens forward pass. Users who relied on ChatGPT left with actionable answers they could not verify. Users who relied on Google left with verified information they could not always act on. Layla's Got You produced neither verified information nor actionable guidance for most users. Producing both requires design systems that detect and respond to the conversion factors present in the target population. No current design accomplished this.

\section{Discussion} Standard HAI evaluation produces tool-level performance averages. CARE produces capability-level profiles disaggregated by conversion factor. The difference is one of unit: usability measures task completion, CARE measures freedom expansion. A tool can score highly on usability while constraining the reproductive freedoms of users whose conversion factors the design does not account for, as all three tools in this study demonstrated. CARE makes the opposing-directions finding visible, that the same design feature expanded capabilities for some users while constraining them for others, by tracing capability outcomes at the individual level rather than aggregating across users. 

\subsection{CARE as a Measurement Layer for HAI} Table \ref{tab:new_eval} maps each Microsoft HAI guideline to the capability question CARE adds above it. Each row represents a shift in what counts as a meaningful evaluation question, from system output to human consequence. The trust row captures this shift most clearly. Traditional evaluation treats high trust as a positive signal of system performance. CARE treats trust as enabling or constraining: high trust in a polished AI suppressed source-checking among participants in this study, constraining Practical Reason and Control over One's Environment in the precise cases where verification mattered most. P3 explained that ChatGPT's confident response convinced them the tool's authority made source verification feel unnecessary. By conventional usability metrics, that interaction succeeded. By CARE, it constrained two capabilities. This reframing produces three evaluation shifts. First, CARE surfaces tradeoffs that system-centered metrics aggregate away: ChatGPT's BLUF structure expanded Practical Reason through actionable responses while the same confident tone constrained it by suppressing verification. Second, CARE makes evaluation user-sensitive: the same tool produced different capability outcomes depending on professional background, gender identity, and prior AI familiarity, differences standard frameworks were built to average rather than disaggregate. Third, CARE makes evaluation appropriate for sensitive domains. In SRH contexts, interaction quality encompasses stigma reduction, emotional support, and bodily autonomy as well as answer accuracy. Layla's empowerment-oriented design expanded Affiliation and Emotion for some participants while its keyword-matching architecture constrained Practical Reason for all, a within-tool tradeoff task completion metrics would not detect. 

\subsubsection{Limitations and Future Work} This study was conducted in an English-language, US-centric context with highly educated participants. SRH conversion factors vary across cultural, linguistic, economic, and legal settings: where SRH is legally restricted, source transparency carries different risks, and in multilingual contexts, jargon operates as a conversion factor differently. Future work should study more diverse populations to measure disparate impact and examine how identity shapes search patterns and trust. We had participants interact with each tool individually to isolate its effects, and screen sharing may have discouraged specific searches or follow-up questions. Future work should examine how people move between tools when available simultaneously, and how a wider variety of tools and questions would build a more comprehensive capability evaluation set.

\subsection{Design Implications} Evaluation should measure capability expansion alongside output quality. All three tools in this study would pass conventional usability and accuracy standards \cite{nielsen2005ten, amershi2019guidelines}, yet CARE revealed capability constraints those standards measure at the task level rather than the freedom level. Designers should evaluate tools against the specific conversion factors present in the target population, since the same design feature produced opposite capability outcomes depending on professional background, gender identity, and prior AI familiarity \cite{selbst2019fairness}. Source visibility is a design decision with capability consequences. Designs that present confident responses without visible citations build implicit trust that suppresses verification, constraining Practical Reason precisely when it matters most. Designers should surface citations by default, giving users source visibility without additional prompting. Citation strategies should account for audience: combining institutional references with community-based resources and plain-language summaries better supports users whose trust in federal health institutions has eroded \cite{steelfisher2023trust}. For users who find sources difficult to interpret, asking generative AI to explain a source in simpler terms offers a partial solution, though this reintroduces risk of inaccurate interpretation. Evaluation validity requires continuous participatory methods with affected populations \cite{wallach2025position, harvey2025understanding}. Researchers should deliberate CARE's capability subset with local communities rather than apply it universally, since conversion factors vary across cultural, linguistic, and legal contexts. Where regulatory infrastructure is limited, capability scorecards comparing tools and community-led auditing offer practical alternatives. 

\subsection{Policy Implications} Current AI safety frameworks \cite{weidinger2022taxonomy, openai2025preparednessframeworkv2} and health regulations like HIPAA address system performance and data privacy without addressing the impact of AI-generated guidance on human freedom. CARE identifies three policy gaps. First, AI systems providing health guidance should cite reputable sources by default, since users in this study frequently could not evaluate the reliability of the advice they received. Second, continuous participatory auditing should precede deployment, with developers publicly reporting capability assessments. Third, policymakers should establish users' rights to know the provenance of health guidance, access underlying sources, and flag information that contradicts clinical experience. North Carolina Senate Bill 963 (2025-2026) \cite{NCBill2026} reflects growing recognition of these gaps, requiring health chatbots to obtain licenses, demonstrate peer-reviewed effectiveness, obtain explicit consent, and impose a duty of loyalty prioritizing users' interests. This provision directly addresses several capability constraints this study identified: absent source transparency and confident tone reduced source verification, constraining Practical Reason and Control over One's Environment. Licensing and safety requirements address system performance and data privacy; capability expansion requires the additional step of testing across the conversion factors present in the target population. Policymakers developing standards for health AI should require evidence that systems were tested across different identities, professional backgrounds, and levels of institutional trust, and should measure effectiveness in terms of capability outcomes, with task completion as one input among several. 

\section{Conclusion}
Standard HAI evaluation measures task performance and interaction quality. We introduced CARE to evaluate capability outcomes as an additional unit of analysis by asking whether tools expand users’ real freedoms. Applying CARE to three SRH tools showed that the same design features expanded capabilities for some users and constrained capabilities for others. Professional background, gender identity, and prior AI familiarity shaped these outcomes beyond the scope of usability and accuracy metrics. Grounded in Sen’s capability approach, CARE operationalizes capability-based evaluation through a Normative Design Lens and an Evaluation Lens. The framework extends to other high-stakes contexts where conversion factors shape capability outcomes.

\subsection{Generative AI Usage Statement}
We evaluate GPT-4o mini in our experiment. We used ChatGPT solely for grammar and language editing of author-written text, in accordance with the conference guidelines.

\bibliographystyle{ACM-Reference-Format}
\bibliography{reference}

\onecolumn 
\appendix
\section*{Appendix}
\vspace{1em}

\section*{Capability Approach Modules \cite{robeyns2017wellbeing}}

\label{app:frameworkModule}

\promptbox[%
    title ={A-Module: The Non-optional Core},
    placeholder = {
\begin{enumerate}
\item[A1.] \textbf{Functionings and capabilities as core concepts:} CARE distinguishes between what users actually achieve (functionings) and their real opportunity to achieve it (capabilities).
\item[A2.] \textbf{Functionings and capabilities are value-neutral categories:} Functionings can be positive (being informed about contraception), negative (receiving stigmatizing responses), or ambiguous (providing care under pressure). Value-neutrality allows CARE to analyze all SRH outcomes. Table \ref{tab:nussbaum_beings_doings} presents the capabilities and functionings selected as critical for reproductive freedom.
\item[A3.] \textbf{Conversion factors:} Personal, social, and environmental factors affect users' ability to transform resources into functionings. CARE introduces sociotechnical conversion factors to account for how tool design mediates this process, explaining why equal access to SRH technologies does not produce equal capabilities.
\item[A4.] \textbf{Means-ends distinction:} Resources are means; capabilities and functionings are ends. In CARE, the SRH technology is the resource, and the end is expanding users' ability to achieve reproductive freedom.
\item[A5.] \textbf{The evaluative space:} SRH technologies must be evaluated by their impact on what users can be and do, not through technical metrics like usage statistics, shifting focus from system performance to human wellbeing.
\item[A6.] \textbf{Other dimensions of ultimate value:} CARE integrates Nussbaum's Ten Central Human Capabilities as its normative foundation.
\item[A7.] \textbf{Valuing each person as an end:} CARE evaluates impact on individual users, ensuring that analysis does not aggregate away disadvantages faced by marginalized populations.
\end{enumerate}
}
]

\vspace{2em}
\promptbox[%
    title = {B-Module: Specification for SRH Technologies},
    placeholder = {
\begin{enumerate}
\item[B1.] \textbf{Purpose:} To provide a framework for normative design and evaluation of SRH technologies that expand users' freedom to access, understand, and act on accurate SRH information free from misinformation, stigma, discrimination, or social sanctions.
\item[B2.] \textbf{Selection of dimensions:} We derived functionings and capabilities by working backward from wellbeing goals. Sexual and reproductive wellbeing encompasses SRH, autonomy and agency, reduced anxiety and harm, and dignity and social inclusion. These goals require key beings (being informed, confident, and respected) and doings (seeking care, making informed choices, and exercising reproductive agency). Drawing on Nussbaum's Ten Central Human Capabilities, we define SRH-related functionings as shown in Table \ref{tab:nussbaum_beings_doings}.
\item[B3.] \textbf{Human diversity:} Users' identities (race, gender, disability) shape their conversion factors and capability sets. Ignoring diversity would overlook how SRH technologies might produce unequal outcomes for users facing compounding barriers.
\item[B4.] \textbf{Agency:} Each user is treated as an independent agent capable of making their own decisions. SRH technologies should support user autonomy by providing options and knowledge rather than prescribing behavior.
\item[B5.] \textbf{Structural constraints:} CARE considers systemic barriers such as inadequate education and SRH stigma that shape interactions between users and technologies, ensuring analysis accounts for the broader structures within which tools operate.
\item[B6.] \textbf{Functionings vs. capabilities:} CARE focuses on expanding capabilities as the precondition for achieving functionings, with reproductive freedom as the ultimate goal.
\item[B7.] \textbf{Meta-theoretical commitment:} CARE adopts a normative approach that centers users from diverse backgrounds and contexts.
\end{enumerate}
}
]
\end{document}